\documentclass[aps,pre,preprint,showpacs,superscriptaddress,noeprint]{revtex4-1}
\usepackage{amsmath,amssymb,physics}
\usepackage{graphicx}
\usepackage{dsfont}
\usepackage{bm}
\usepackage{xcolor}
\usepackage{microtype}
\usepackage[colorlinks=true,linkcolor=blue,citecolor=blue,urlcolor=blue]{hyperref}
\usepackage[all]{hypcap} 

\def\rms{\rm\scriptscriptstyle}

\def\dd{{\rm d}}

\begin{document}

\title[Collective excitations in jammed states]{Collective excitations 
in jammed states: ultrafast defect propagation and finite-size scaling}

\author{Alexander P.\ Antonov}
\affiliation{Universit{\" a}t Osnabr{\" u}ck, Fachbereich Physik,
  Barbarastra{\ss}e 7, D-49076 Osnabr{\" u}ck, Germany}

\author{David Vor\'a\v{c}}
\affiliation{Charles University, Faculty of Mathematics and Physics,
  Department of Macromolecular Physics, V Hole\v{s}ovi\v{c}k\'ach 2,
  CZ-18000 Praha 8, Czech Republic}

\author{Artem Ryabov}
\affiliation{Charles University, Faculty of Mathematics and Physics,
  Department of Macromolecular Physics, V Hole\v{s}ovi\v{c}k\'ach 2,
  CZ-18000 Praha 8, Czech Republic}

\author{Philipp Maass}
\affiliation{Universit{\" a}t Osnabr{\" u}ck, Fachbereich Physik,
  Barbarastra{\ss}e 7, D-49076 Osnabr{\" u}ck, Germany}

\date{November 15, 2022} 

\begin{abstract}
In crowded systems, particle currents can be mediated by propagating collective excitations which are generated as rare events, are localized
and have a finite lifetime. The theoretical description of such excitations is hampered by the problem of identifying
complex many-particle transition states, calculation of their free energies, and the evaluation of propagation mechanisms and velocities. Here we show that these problems can be tackled for a highly jammed system of hard spheres in a periodic potential.
We derive generation rates of collective excitations, their anomalously high velocities, explain the occurrence of an apparent jamming transition and its strong dependence on the system size. The particle currents follow a scaling behavior, where
for  small systems the current is proportional to the generation rate and for large systems given by the geometric mean of the 
generation rate and velocity. Our theoretical approach is widely applicable to dense nonequilibrium systems in confined geometries. It provides new perspectives for studying dynamics of collective excitations in experiments.
\end{abstract}

\maketitle  
\newpage 

\section{Introduction}
\label{sec:introduction}
In molecular transport at the nanoscale, spatial confinements slow down particle motion that ultimately can become arrested in a jammed state. The slowing down is in particular observed 
at high particle 
densities in single-file dynamics, where particles cannot overtake each other \cite{Wei/etal:2000, Taloni/etal:2017}. 
Such transport ubiquitously occurs in biological traffic \cite{Schadschneider/etal:2010, Chou/etal:2011},
like in the protein synthesis by ribosomes \cite{MacDonald/etal:1968} and
molecular motor motion along filaments \cite{Kolomeisky:2013, Appert-Rolland/etal:2015}.
It is highly relevant also for nanotechnology and chemical engineering including catalytic processes in zeolites \cite{Hahn/etal:1996}, molecular sieves, as well as for control and steering of flows in nanotubes \cite{Cheng/Bowers:2007, Dvoyashkin/etal:2014}, and membrane channels and pores \cite{Bauer/Nadler:2006,  Kahms/etal:2009, Bressloff/Newby:2013}. 

Paradigmatic models for biological single-file transport are the asymmetric simple exclusion process (ASEP) and its variants \cite{Derrida:1998, Schuetz:2001, Blythe/Evans:2007, Chowdhury:2013, Pillay/etal:2018, Riba/etal:2019, Mines/etal:2022}. In ASEP \cite{Derrida:1998, Schuetz:2001}, a particle current is induced by letting particles hop between neighboring sites of a one-dimensional lattice with a bias in one 
direction. Each lattice site can be occupied by at most one particle, reflecting hardcore (steric) interactions. 
Over the last decades, this simple process has become a fundamental model for nonequilibrium statistical mechanics. It provides one of the rare cases, where the statistical mechanics of nonequilibrium steady states can be 
derived exactly. Thanks to these exact results, many intriguing phenomena could be understood on a firm basis.
Most prominent is perhaps the occurrence of phase transitions, which cannot appear in equilibrium one-dimensional systems. These phase transitions are related to the nonlinear dependence of the current 
on the particle density \cite{Krug:1991, Antal/Schuetz:2000, Dierl/etal:2013}.  In ASEP, 
the nonlinear current-density relation is a direct consequence of
particle hops becoming increasingly blocked with higher density. This eventually leads to jamming when the number of particles equals the number of lattice sites (filling factor one).

For single-file transport in zeolites, membrane channels and pores, the coarse-grained lattice description is less appropriate. 
Recently, we have shown that the dynamics
in these systems should be very different from ASEP-like behavior \cite{Lips/etal:2018, Lips/etal:2019, Antonov/etal:2021}.
Inspired by ASEP, we suggested the Brownian asymmetric exclusion process (BASEP) \cite{Lips/etal:2018} to model driven 
single-file motion in periodic potentials. In BASEP, hard-sphere interacting particles with diameter $\sigma$ are driven by a 
constant external force $f$ in a cosine potential with wavelength $\lambda$. 
In connection with a coarse-grained lattice description, one may think of 
the potential wells to represent the lattice sites and the force $f$ to give rise to preferential hopping of particles between 
neighboring wells. However, while the wavelength $\lambda$ of the periodic potential plays the role of a lattice constant, 
the particle size $\sigma$ turns out to be a very important length scale that leads to novel collective effects that are absent in a 
corresponding lattice description. 

The BASEP is useful to interpret nonequilibrium phenomena studied experimentally in soft matter and microfluidic systems, where 
the particle size has a strong influence on the dynamics. It shows various interesting features, as many-particle induced current 
enhancement with respect to noninteracting 
particles, current suppression due to mutual particle blocking, and commensurability effects when $\sigma$ becomes close to 
multiples of the wavelength $\lambda$ of the periodic potential.  
In recent experiments, the BASEP was helpful to understand fundamental differences between flow- and force-driven single-file 
transport of colloids in fluidic environments \cite{Cereceda-Lopez/etal:2021}.

\begin{figure}[t!]
\center{\includegraphics[width=0.6\textwidth]{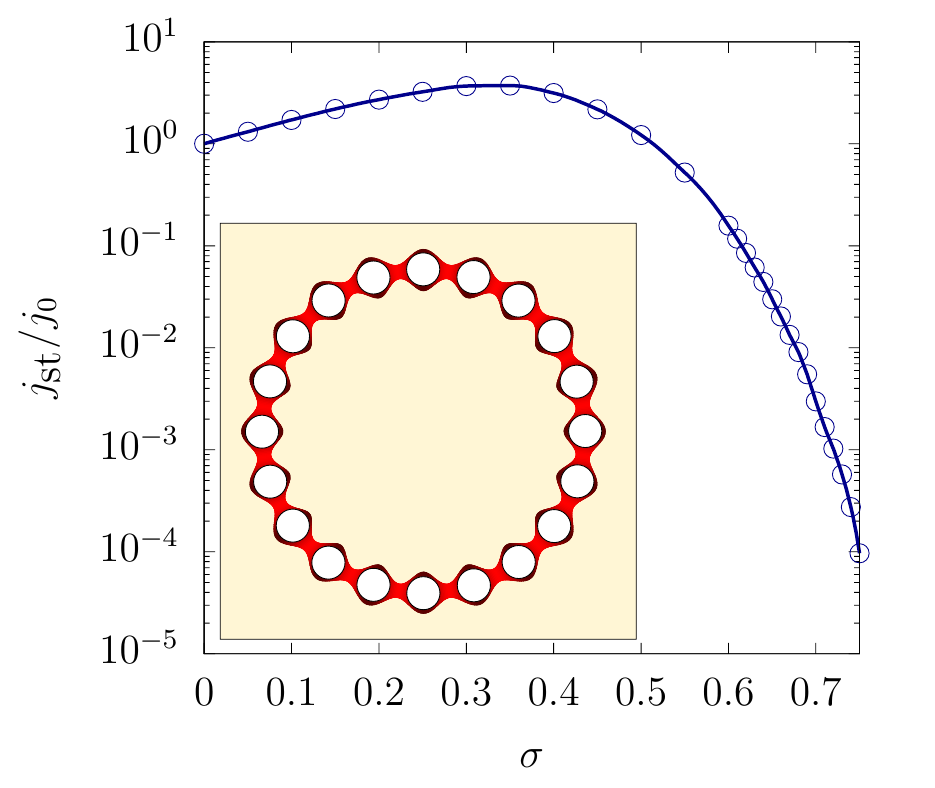}}
\caption{Normalized steady-state current $j_{\rm st}/j_0$ for $N=20$ hard spheres as a function of their diameter $\sigma$. The spheres are performing an overdamped Brownian motion in the cosine potential $U(x)=(U_0/2)\cos(2\pi x/\lambda)$ with $N$ potential wells (filling factor one) subject to periodic boundary conditions. They are dragged by a constant force $f$. The normalization $j_0$ refers to the current of noninteracting particles. Symbols mark results from Brownian dynamics simulations and the line is drawn as a guide for the eye. Parameters are $U_0=6$ and $f=0.2$ in the chosen units 
($k_{\rm B}T=1$, $\lambda=1$). The inset illustrates an experimental realization of the model \cite{Cereceda-Lopez/etal:2021}:
20~microparticles are confined to 20 optical traps (potential wells). 
Particle transport along the ring becomes possible after creating an empty potential well (hole) by a thermal excitation.}
\label{fig:j-sigma}
\end{figure}

Here, we show that an apparent jamming transition occurs in BASEP for  $\rho=1$, when $\sigma$ approaches values of about $0.75\lambda$. This transition is demonstrated in Fig.~\ref{fig:j-sigma} for a periodic
system with 20 potential wells. It manifests itself in a current decreasing by several orders of magnitude when $\sigma$ is 
increased from 0.6 to $0.75\lambda$. An experimental realization of this setting \cite{Cereceda-Lopez/etal:2021} is sketched in the
the inset of Fig.~\ref{fig:j-sigma}. In the experiments, the current would drop essentially to zero, like in a jamming transition with complete arrest of particle flow. 

Considering the complete filling of all the potential wells, it may appear surprising 
that particle flow can still be present at large particle diameters $\sigma$. However, transport in the apparently jammed phase is still 
possible by cluster-like defects with a certain lifetime that propagate in the direction of the drag force. 
We refer to them as collective excitations \cite{Scalliet/etal:2019}, which on entropic 
reasons must occur in any crowded system, albeit rarely. 

These localized excitations are different from low-frequency 
wave-like collective excitations, which also occur in hard-sphere systems \cite{Huerta/etal:2021}.
They are defects composed of several particles and are generated via well-defined transition states by overcoming free energy barriers 
that depend on the particle 
size $\sigma$. After formation, the defects propagate with surprisingly high velocity. The defect lifetimes depend strongly on the 
system size $L$,  which gives rise to pronounced variation of the defect-mediated currents with $L$. In particular,
the apparent jamming transition  becomes less pronounced with increasing $L$. The sensitive dependence of the currents on both 
$L$ and $\sigma$ follows a scaling law. 

Generally, it is very difficult to gain a thorough understanding of the generation and motion of collective excitations in crowded
systems, because of the complex particle configurations involved in these processes. Also, it is hard to study collective excitations 
in experiments. The system in Fig.~\ref{fig:j-sigma} is suitable to meet both these
theoretical and experimental challenges.


\section{Defect-mediated transport}
\label{sec:defect-transport}

The Brownian dynamics of $N$ particles in BASEP is given by the Langevin equations 
\begin{equation}
\frac{\dd x_i}{\dd t}=\mu\left(f -U'(x_i)\right) + \sqrt{2D}\,\eta_i(t)\,,\hspace{1em}i=1,\ldots,N\,,
\label{eq:langevin}
\end{equation}
where $\mu$ is the particle mobility, $D=k_{\rms{B}}T\mu$ is the bare diffusion coefficient of a single particle in a flat potential
with $k_{\rms B}T$  the thermal energy, and $\eta_i(t)$ are Gaussian 
white noise processes with  zero mean and correlations $\langle \eta_i(t) \eta_j(t') \rangle = \delta_{ij}\delta(t - t')$.
The function
\begin{equation}
U(x) = \frac{U_0}{2} \cos \left(\frac{2 \pi x}{\lambda} \right)
\label{eq:cosine-potential}
\end{equation}
is the external $\lambda$-periodic potential, where $U_0\gg k_{\rms{B}}T$, and the constant force $f$ is driving the particles.
Due to the hard-sphere interaction,
the distance between neighboring particles cannot be smaller than $\sigma$
and the particles keep their order (single-file transport). The filling factor of the potential wells is $\rho=N\lambda/L$, 
where $L$ is the system length. 

Several fast simulation algorithms have been developed in the past to tackle the hard-sphere interactions in Brownian dynamics simulations 
\cite{Tao/etal:2006, Scala/etal:2007, Scala:2012, Behringer/Eichhorn:2012, Sammueller/Schmidt:2021}.  
When we refer to simulation results in the following, we always used the algorithm given by Scala et al.\ \cite{Scala/etal:2007}. This is an
event-driven scheme
\footnote{Encounters between two particles in small time steps $\Delta t$
are treated as elastic collisions. Let $ \Delta x_i$ be the attempted displacement
of a particle $i$ in the time interval $\Delta t$ according to Eq.~(\ref{eq:langevin}) when ignoring hard-sphere interactions.
If that $\Delta x_i$ in connection with the $\Delta x_j$ of its neighboring particles does not lead to an encounter of the respective particles, 
the particle $i$ is displaced accordingly. If particle $i$ would encounter particle $j$, they are treated as freely 
propagating hard spheres exchanging their fictive velocities 
$\Delta x_i / \Delta t$ and $\Delta x_j / \Delta t$, 
i.e.\ the displacement of particle $i$ is the result of an elastic collision. If there is a subsequent collision, it is treated analogously.}
and details of our implementation can be found in Ref.~\cite{Ryabov/etal:2019}.
A new approach to this problem, which addresses
the theoretical basis of a treatment in arbitrary external force fields,
will be given in Ref.~\cite{Antonov/etal:2022-tobepublished}.

We applied periodic boundary conditions for a system size $L$ being an integer multiple of $\lambda$.
As units of energy, length and time, we used $k_{\rms B}T$, $\lambda$ and $\lambda^2/D$, respectively. 
The potential amplitude is fixed to 
$U_0=6$ and this value much larger than the thermal energy implies that a single-particle motion 
in the periodic potential would be a 
thermally activated hopping on the scale of the Kramers' time. The drag force is fixed to 0.2, i.e., $f\lambda/k_{\rms B}T=0.2$,
meaning that we consider a weak driving 
much below the critical ``tilting force'' $f_c=\pi U_0/\lambda\cong18.8$, above which potential minima would no longer be present.

The apparent jamming transition in Fig.~\ref{fig:j-sigma} occurs for filling factor $\rho=1$ and is reflected in a rapid decrease of the steady-state current $j_{\rm st}$ with growing particle size when $\sigma$ becomes larger than 0.6.
For small particle diameters, $j_{\rm st}$ increases because of
a barrier reduction effect discussed earlier \cite{Lips/etal:2018, Vorac/etal:2020, Antonov/etal:2021}. Here we focus our study on the regime $\sigma>0.6$ and the defect-mediated transport associated with the apparent jamming transition. 
In the following, we characterize the defects and
calculate defect generation rates and lifetimes, which in turn enable us to predict transport properties.

\begin{figure}[t!]
\includegraphics[width=\textwidth]{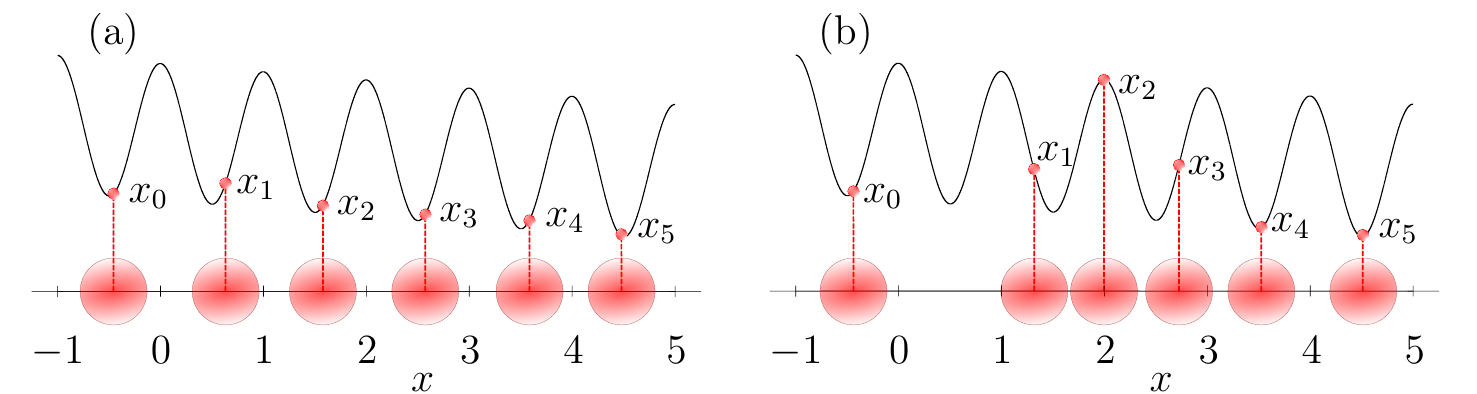}
\caption{Sketch of local configurations of six hard spheres with diameter $\sigma=0.65$ in the tilted potential $U(x)-fx$. In 
(a) a typical state is shown, where the particles are located close to positions of mechanical equilibria. These
positions are at $m+1/2+\xi$, $m=-1,0,\ldots$, where $\xi$ is given in Eq.~(\ref{eq:xi}) and has the
small value $\xi\cong0.017$ for our parameters $U_0=6$ and $f=0.2$. In (b) a rare transition state is depicted, where a hole is present around the minimum at $1/2+\xi$ (empty well between maxima at $-\xi$ and $1-\xi$). The transition state
is formed by two particles (2-particle transition state): 
the first particle to the right of the hole has not returned to the empty potential well, while the 
the second particle is located at the maximum at $2-\xi$.}
\label{fig:sketch_configurations}
\end{figure}


\begin{figure}[b!]
\centering\includegraphics[width=\textwidth]{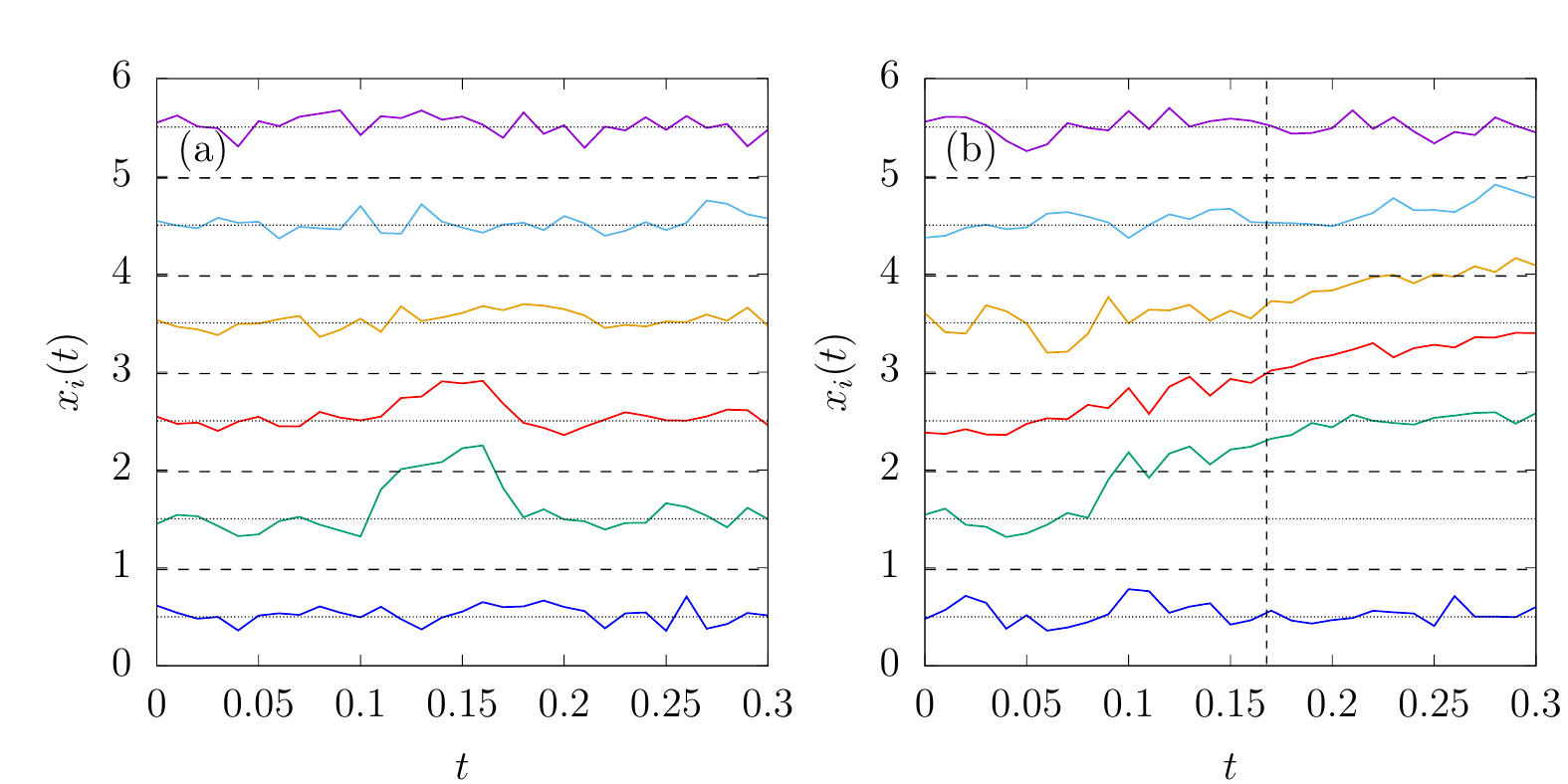}
\caption{Simulated particle trajectories for $\sigma=0.65$ that illustrate
possible system evolutions right after a hole formation. Horizontal dotted 
and dashed lines mark minima and maxima of the tilted potential, respectively.
In (a) a hole appears at time $t\simeq 0.12$ and is refilled at $t\simeq 0.17$ (see trajectory marked in green). 
In (b) a 2-particle transition state for a propagating defect forms at a time  
indicated by the vertical dashed line. This transition state corresponds to the one depicted schematically in Fig.~\ref{fig:sketch_configurations}(b). The propagation of the defect formed after its generation is displayed
in Fig.~\ref{fig:trajectories-propagation}(a).}
\label{fig:defect-formation-in-simulation}
\end{figure}

\subsection{Defect generation rate}
\label{subsec:defect-generation-rate}
In a system with filling factor $\rho=1$, the energetically preferred particle configuration would be the crystalline type shown in Fig.~\ref{fig:sketch_configurations}(a): all particles are residing close to the minima of the tilted periodic potential $U(x)-fx$. 
The minima and maxima of the tilted potential are at positions $m-\xi$ and $m+1/2+\xi$ with $m$ an integer, respectively, where
\begin{equation}
\xi=\frac{1}{2\pi}\arcsin\Bigl(\frac{\pi f}{U_0}\Bigr)\,.
\label{eq:xi}
\end{equation}
If the particles have a
diameter $\sigma>0.5$, double occupancies of potential wells are highly improbable. At 
first sight, the system appears to be frozen and it is 
difficult to imagine how a weak driving force can generate a current. However, thermal 
excitations can lead to defects, where a particle escapes a potential well and leaves an unoccupied well behind, as 
illustrated for particle~1 in Fig.~\ref{fig:sketch_configurations}(b). We call an unoccupied potential well a hole.

After hole formation, several particles form a defect next to the hole. 
This defect can disappear after a short transient time, if the particle
that has escaped the potential well and left the hole behind, returns 
to the well, see the particle trajectory marked in green in
Fig.~\ref{fig:defect-formation-in-simulation}(a). More interesting is the case where 
the defect starts to propagate in the bias direction, see Fig.~\ref{fig:defect-formation-in-simulation}(b). 
On average over time, the hole then moves slowly against the drag force due to a biased hopping-like 
motion of the particles next to it. 

\begin{figure}[t!]
\includegraphics[width=\textwidth]{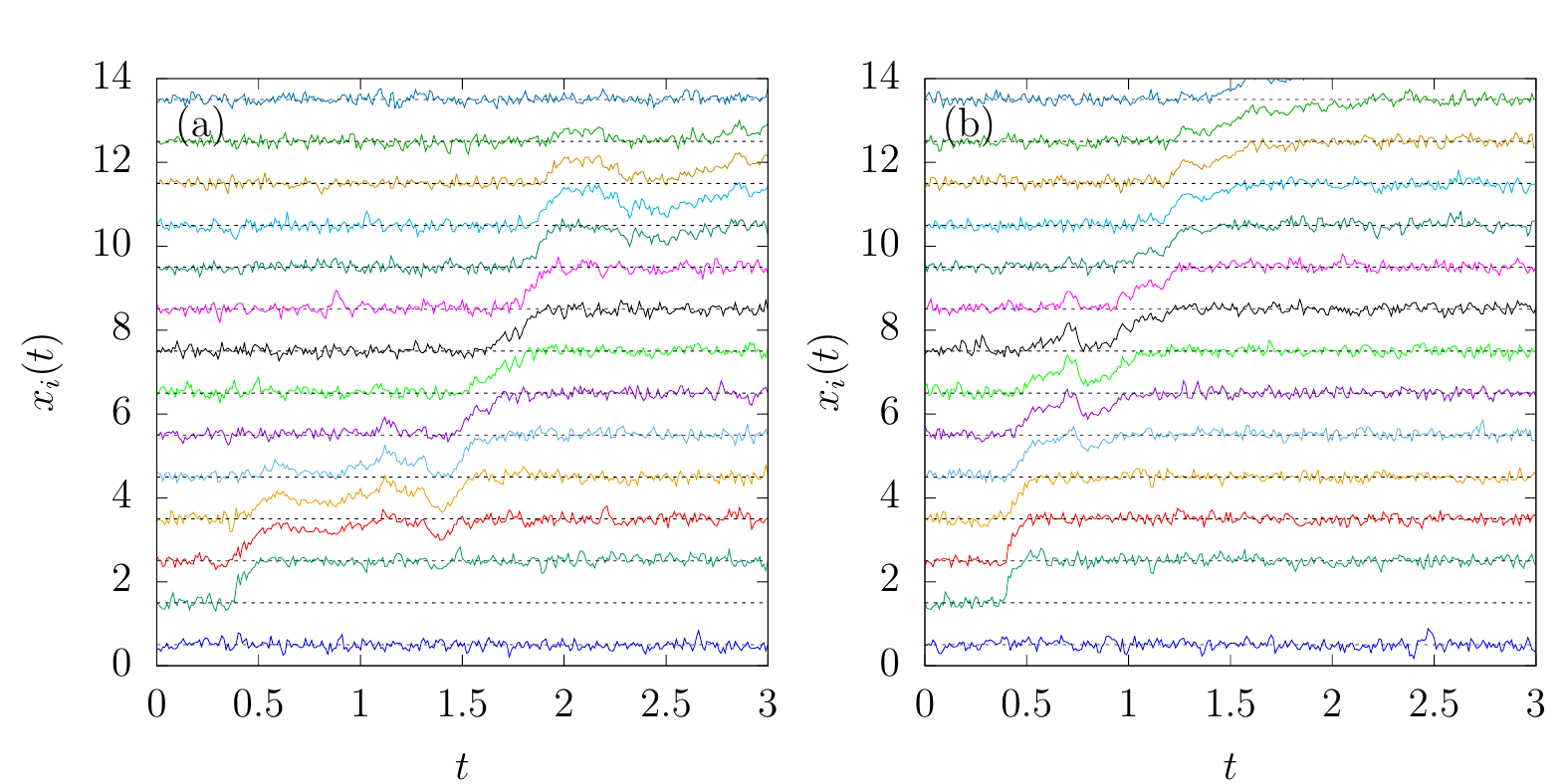}
\caption{Propagating defects for (a) $\sigma=0.65$ and (b) $\sigma=0.75$.  In (a) the defect is generated via the 2-particle transition state displayed in Fig.~\ref{fig:defect-formation-in-simulation}(b), while for the larger particle diameter in (b) the generation of the defect is via a 3-particle transition state at time $t\simeq0.5$. The propagating defect in (b) involves a larger number of particles and moves faster than the one in (a).}
\label{fig:trajectories-propagation}
\end{figure}

The reason for the different time evolutions after hole formation is that in Fig.~\ref{fig:sketch_configurations}(b) a transition state
for a propagating defect is generated, while this is not the case in Fig.~\ref{fig:sketch_configurations}(a).
In the transition state, the second particle to the right of the hole [red trajectory in Fig.~\ref{fig:sketch_configurations}(b)] 
reaches the second potential maximum from the hole in bias direction (position 3 in the figure), while the hole is still present. The 
time instant, when this happens, is indicated by the vertical line. We found that such event is a prerequisite for obtaining a 
propagating defect shown in Fig.~\ref{fig:trajectories-propagation}(a). 

We consider a propagating defect generated by two 
``excited particles'' as a one with a 2-particle transition state.
For larger particle diameters, also $n$-particle transition states with $n=3,4,\ldots$ can occur. In these transition states,
particle~$n$ must reach the potential 
maximum at a distance of $1/2+(n-1)$ wavelengths from the hole, and particles $1,\ldots,(n-1)$ all need to have
passed one potential barrier. If the hole is located at $x=1/2+\xi$ as in Fig.~\ref{fig:sketch_configurations}(b), the coordinates  $x_1,\ldots,x_{n-1}$ of these particles must satisfy the conditions $x_i>(i-\xi)$, $i=1,\ldots,(n-1)$. This defines the 
$n$-particle transition state for generating a propagating defect.



For calculating the free energy $F_n^\ddagger(\sigma)$ of the transition state and the free energy $F_n^0(\sigma)$
of a corresponding reference state, 
we consider two boundary particles placed at fixed positions of mechanical equilibrium.
To be specific, the boundary particle to the left is at position $x_{\rm b}^-=-1/2+\xi$ and the boundary particle to the right
at position $x_{\rm b}^+=x_{\rm b}^-+2n+1$. This implies that we place these boundary particles at a distance of $(2n+1)$ 
wavelengths, which allows $2n$ particles to occupy $2n$ different potential wells in the reference state. 
We found that it is sufficient to consider $n$ additional particles to the right of the particle $n$ at the potential maximum, because
taking a larger number is not affecting the results for $F_n^\ddagger(\sigma)$ and $F_n^0(\sigma)$. 

In the transition state, the particle $n$ must be at the position $x_n^\ddagger=n-\xi$.
Therefore, the position of the particle $n$ is also fixed in the reference state, but now at the position of mechanical equilibrium at $x_n^0=n-1/2+\xi$. The free energy of the reference state
then is $F_n^0(\sigma)=-\ln Z_n^0(\sigma)$ with the partition sum
\begin{align}
Z_n^0(\sigma)&=\exp\bigl\{-[U(x_n^0)-fx_n^0]\bigr\}\hspace{0em}
\int\limits_{x_n^0+\sigma}^{x_{\rm b}^+-n\sigma}\hspace{-0.7em}\dd x_{n+1}\hspace{-1em}
\int\limits_{x_{n+1}+\sigma}^{x_{\rm b}^+-(n-1)\sigma}\hspace{-1.5em}\dd x_{n+2}\hspace{0.1em}\ldots\hspace{-1.1em}
\int\limits_{x_{2n-1}+\sigma}^{x_{\rm b}^+-\sigma}\hspace{-1em}\dd x_{2n}
\label{eq:Z0}\\
&\hspace{6em}{}\times\hspace{-1em}
\int\limits_{x_{\rm b}^-+(n-1)\sigma}^{x_n^0-\sigma}\hspace{-1em}\dd x_{n-1}\hspace{-1em}
\int\limits_{x_{\rm b}^-+(n-2)\sigma}^{x_{n-1}-\sigma}\hspace{-1em}\dd x_{n-2}\hspace{0.1em}\ldots\hspace{-0.5em}
\int\limits_{x_{\rm b}^-+2\sigma}^{x_3-\sigma}\hspace{-0.5em}\dd x_2\hspace{0.1em}\hspace{-0.5em}
\int\limits_{x_{\rm b}^-+\sigma}^{x_2-\sigma}\hspace{-0.5em}\dd x_1
\exp\Biggl\{-\hspace{-0.5em}\sum_{j=1,j\ne n}^{2n} [U(x_j)-fx_j]\Biggr\}.\nonumber
\end{align}

In the transition state, in addition to setting $x_n^\ddagger=n-\xi$, we must require the potential well covering the
interval $[-\xi,1-\xi]$ to be empty (representing a hole).
The first $(n-1)$ particles 
occupy positions in the interval $[j-\xi,x_{j+1}-\sigma]$, $j=1,\ldots, (n\!-\!1)$,
and the remaining $n$ particles occupy positions in the interval
$[x_{j-1}+\sigma,x_{\rm b}^+-(2n+1-j)\sigma]$, $j=(n\!+\!1),\ldots, 2n$. 
Accordingly, the free energy of the transition state is $F_n^\ddagger(\sigma)=-\ln Z_n^\ddagger(\sigma)$ with
\begin{align}
Z_n^\ddagger(\sigma)&=\exp\bigl\{-[U(x_n^\ddagger)-fx_n^\ddagger]\bigr\}\hspace{0em}
\int\limits_{x_n^\ddagger+\sigma}^{x_{\rm b}^+-n\sigma}\hspace{-0.7em}\dd x_{n+1}\hspace{-1em}
\int\limits_{x_{n+1}+\sigma}^{x_{\rm b}^+-(n-1)\sigma}\hspace{-1.5em}\dd x_{n+2}\hspace{0.1em}\ldots\hspace{-1.1em}
\int\limits_{x_{2n-1}+\sigma}^{x_{\rm b}^+-\sigma}\hspace{-1em}\dd x_{2n}
\label{eq:Zddagger}\\
&\hspace{8em}{}\times\hspace{0em}
\int\limits_{n-1-\xi}^{x_n^\ddagger-\sigma}\hspace{-0.5em}\dd x_{n-1}\hspace{-1em}
\int\limits_{n-2-\xi}^{x_{n-1}-\sigma}\hspace{-0.5em}\dd x_{n-2}\hspace{0.1em}\ldots\hspace{-0.5em}
\int\limits_{2-\xi}^{x_3-\sigma}\hspace{-0.5em}\dd x_2\hspace{0.1em}\hspace{-0.5em}
\int\limits_{1-\xi}^{x_2-\sigma}\hspace{-0.5em}\dd x_1
\exp\Biggl\{-\hspace{-0.5em}\sum_{j=1,j\ne n}^{2n} [U(x_j)-fx_j]\Biggr\}.\nonumber
\end{align}

The free energy barrier between the
transition and reference state yields the rate $\lambda_n(\sigma)$ for generating a propagating defect 
via an $n$-particle transition state, 
\begin{equation}
\lambda_n(\sigma)=\nu\exp[-(F_n^\ddagger(\sigma)-F_n^0(\sigma))]
=\nu\,\frac{Z_n^\ddagger(\sigma)}{Z_n^0(\sigma)}\,.
\label{eq:lambdan}
\end{equation}
Here, $\nu$ is a bare rate, which should be of the order of $D/\lambda^2$, i.e., one in our units. 

\begin{figure}[t!]
\centering\includegraphics[width=0.55\textwidth]{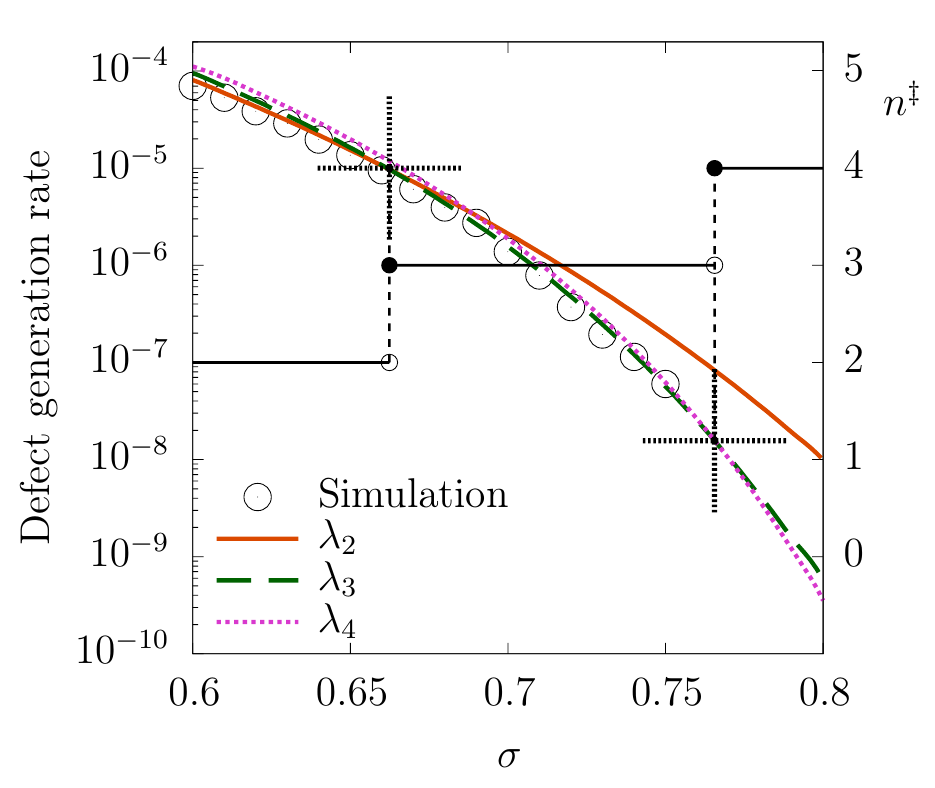}
\caption{
Defect generation rates (left axis) for different particle diameters $\sigma$. The rates $\lambda_n$ were calculated from Eq.~(\ref{eq:lambdan}). The number $n^\ddagger$ of particles involved in
the transition state (right axis) increases with $\sigma$ in a staircase-like manner.
The steps of this staircase are determined by Eq.~(\ref{eq:nd}) and
occur at particle diameters, where lines for different $\lambda_n$ intersect, as indicated by the crosses.}
\label{fig:generation-rate}
\end{figure}

Figure~\ref{fig:generation-rate} (left axis) shows $\lambda_2$, $\lambda_3$ and $\lambda_4$ 
as functions of $\sigma$ in comparison with simulated data, where we performed 
the integrations in Eqs.~(\ref{eq:Z0}) and (\ref{eq:Zddagger}) 
numerically and set $\nu=1$. In the simulations, we determined $\lambda_{\rm gen}$ by counting the mean 
number $\bar N_{\rm def}$ of propagating defects in 
a large time interval $\Delta t$ in the non-equilibrium steady state, yielding $\lambda_{\rm gen}=\bar N_{\rm def}/\Delta t$.
The data are plotted for the range $\sigma\in[0.6,0.75]$, where we could determine 
the generation rate $\lambda_{\rm gen}$ of propagating defects with reliable accuracy.
Note that the formation of a defect is a rare thermally activated event and that $\lambda_{\rm gen}$ decreases 
faster than exponentially with $\sigma$. In the considered $\sigma$-interval, $\lambda_{\rm gen}$ falls
by about three orders of magnitude, reaching extremely small values for $\sigma\gtrsim0.7$. 

The simulated data of $\lambda_{\rm gen}$ in Fig.~\ref{fig:generation-rate} agree well with the calculated $\lambda_2$ for 
$\sigma\lesssim0.66$ and with the calculated $\lambda_3$ for $\sigma\gtrsim0.66$. In fact, we found that the smaller of the two calculated rates gives a very good prediction of $\lambda_{\rm gen}$, i.e.,
\begin{equation}
\lambda_{\rm gen}(\sigma)\cong\min_{n}\{\lambda_n(\sigma)\}\,.
\end{equation}
This can be formulated as a maximum free energy barrier principle: for given $\sigma$, the particle number $n^\ddagger$
defining the transition state follows 
by maximizing the free energy barrier $\Delta F^\ddagger_n(\sigma)=[F_n^\ddagger(\sigma)-F_n^0(\sigma)]$ with respect to $n$,
\begin{equation}
n^\ddagger(\sigma)={\rm arg\hspace{-0.2em}}\max\limits_{\hspace{-1.2em}n}\{\Delta F^\ddagger_n(\sigma)\}\,.
\label{eq:nd}
\end{equation}
Its generation rate then is $\lambda_{\rm gen}=\lambda_{n^\ddagger}$. The staircase-like increase of
$n^\ddagger$ with the particle diameter
predicted by the maximum free energy barrier principle is displayed in Fig.~\ref{fig:generation-rate} (right axis).
The change from the 2-particle to the 3-particle transition state should occur at $\sigma\cong0.66$.
The examples of defects generation and propagation shown in Fig.~\ref{fig:trajectories-propagation} are in agreement 
with this prediction.

One may ask why a principle of maximum instead of minimum free energy barrier is present. This
has the following explanation. First, we note that to create an $n$-particle transition state, the $(n-k)$-particle transition states with
$k=1,\ldots,(n\!-\!2)$ need to be created as intermediate states. Second, we point out that
for creating a propagating defect, the hole must not refill, i.e., a stabilized hole needs to be created. For smaller $\sigma$, it is 
enough to have one single-occupied well to the right of the hole for stabilizing it. 
This explains why for $0.6<\sigma\lesssim0.66$ in 
Fig.~\ref{fig:generation-rate} the rate $\lambda_2$ is giving $\lambda_{\rm gen}$: the 2-particle transition state must be created
first and it is sufficient to obtain a stabilized hole and a propagating defect.
 When increasing $\sigma$, the 
particles in the double-occupied next-neighboring well to the right have a higher tendency to push the particle from the
neighboring well back to the hole. The hole therefore becomes stabilized only after two single-occupied wells are created right of it. This leads to the three-particle transition state as the relevant one for generating a propagating defect and explains why 
the rate $\lambda_3$ is giving $\lambda_{\rm gen}$ for $0.66\lesssim\sigma<0.75$: the 2-particle transition state must be created
first, but it is not sufficient for obtaining a stabilized hole, as the two particles in the double-occupied next-neighboring well of the hole are pushing the particle back into the next-neighboring well. When
increasing $\sigma$ further, we conjecture that this mechanism will proceed: the number of single-occupied wells needed to stabilize the hole increases. This protects the hole against refilling from the right.
We could not confirm this mechanism for $\sigma>0.75$ in the simulations as the events of propagating 
defect generation become too rare.

The very strong decrease of $\lambda_{\rm gen}$ with $\sigma$ in Fig.~\ref{fig:generation-rate} suggests that it is responsible for the apparent jamming transition seen in Fig.~\ref{fig:j-sigma}. However, the defect-mediated current is not only dependent on the
generation rate of the defects but also on the defect velocity. To obtain a complete picture we thus need to understand the defect velocities and their dependence on $\sigma$.

\begin{figure}[t!]
\centering\includegraphics[width=0.55\textwidth]{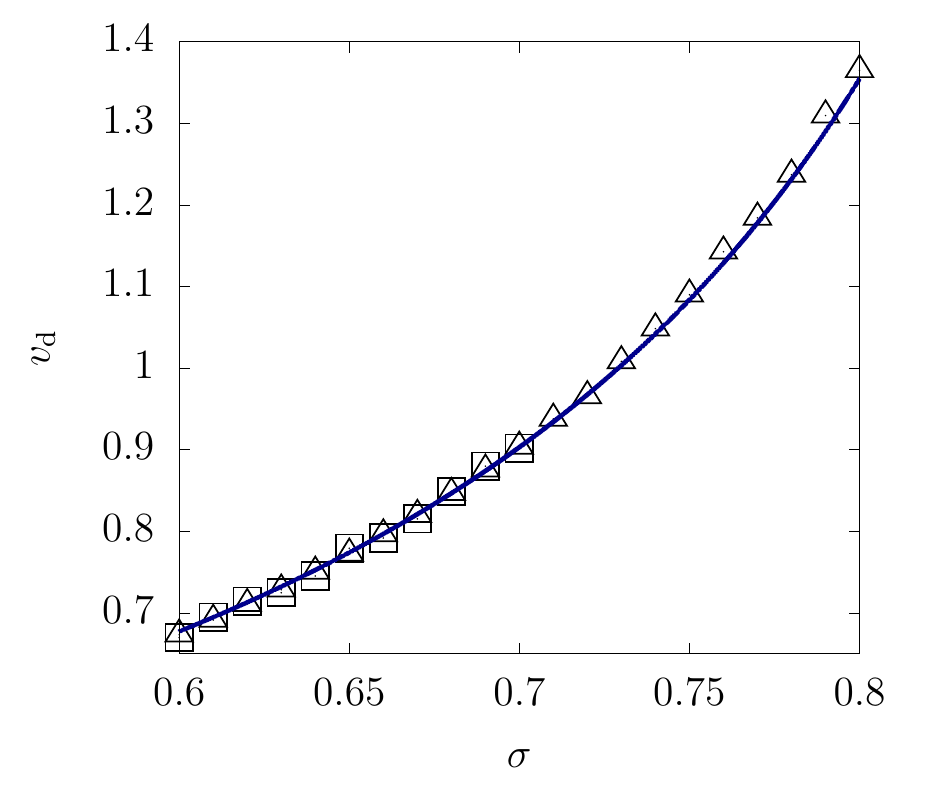}
\caption{Velocity of propagating defects in dependence of the particle diameter. The squares show simulation results for
a system with filling factor one, and the triangles represent the results from the inserted particle method (11~particles 
in a system with 10 potential wells). The line refers to the theoretical prediction given by Eq.~(\ref{eq:vd}) with
$\alpha=1.35$.}
\label{fig:defect-velocity}
\end{figure}

\subsection{Defect velocity}
\label{subsec:velocity}
To calculate the defect velocity $v_{\rm d}$ in simulations, we recorded the time instant $t_{\rm gen}$ when a 
propagating defect is generated and the time instant $t_{\rm ann}$  
when this defect becomes annihilated due to recombination with a hole. 
In the time interval $[t_{\rm gen},t_{\rm ann}]$,
we followed the trajectory of the double-occupied well in the defect. Then we excluded the initial 
period $[t_{\rm gen},t_{\rm gen}+1]$ because right after the defect generation, the propagation of the double-occupied well
is much faster than on average, see also the defect initiation periods 
in Figs.~\ref{fig:trajectories-propagation}(a) and~(b). A similar effect occurs before the recombination with a hole 
and we therefore excluded also the time interval 
$[t_{\rm ann}-1,t_{\rm ann}]$ from the analysis. 
The defect velocity is obtained from the distance traveled by the double-occupied well within the 
considered time interval.

In this method, an averaging of $v_{\rm d}$ over many individual defect trajectories 
requires a high computational effort because of the low defect generation rate, 
in particular at larger $\sigma$. A more convenient method is to create a defect artificially by inserting an additional
particle into the system with filling factor one. This immediately creates one defect and allows to calculate $v_{\rm d}$ from mean travelled distances per time. Results from both methods agree, see Fig.~\ref{fig:defect-velocity}. In principle, it would be possible to create defects also by inserting more than one particle into the system with filling factor one. In that case, however, more than one defect can form. 
This does not only complicate the determination of the velocities of each defect but the defect velocities can then be influenced also by effective defect-defect interactions.

The dependence of $v_{\rm d}$ on $\sigma$ in Fig.~\ref{fig:defect-velocity} is very weak compared to the 
change of $\lambda_{\rm gen}$ in Fig.~\ref{fig:generation-rate}, note the different scales of the vertical axes. 
While $\lambda_{\rm gen}$ decreases by more than three orders of magnitude when $\sigma$ is increased from 0.6 to 0.75, the 
defect velocity increases by a factor of about 1.5 only. What is remarkable is the magnitude of $v_{\rm d}$, which is about
hundred times larger than the velocity $v_0\cong8.4\times10^{-3}$ of noninteracting particles (this can be calculated based on an exact result for a single particle, see \cite{Stratonovich:1958, Ambegaokar/Halperin:1969}). It is even about four times larger than the velocity $f=0.2$ ($\mu=1$ for the chosen units) of independent particles in a flat potential.

To understand the behavior of $v_{\rm d}$, let us note that during the defect motion compact cluster-like structures appear where four or more particles are nearly in contact with each other, see Figs.~\ref{fig:trajectories-propagation}(a) and (b).
This motivates the following rough estimation: let us consider a cluster of $(n+1)$ particles in contact with each other, where the first (leftmost) particle is at the minimum of the tilted potential and the remaining $n$ particles are moving as a compact cluster until the second particle reaches a point of mechanical equilibrium. To reach this point, the cluster has to move a distance $(1-\sigma)$.
After reaching it, the whole defect is shifted by one wavelength, because the second particle is now adopting the role of the first particle. Hence we can say that the defect velocity is $v_{\rm d}(\sigma)=1/t_\sigma$, where $t_\sigma$ is the mean time of the cluster to move a distance $(1-\sigma)$.

The cluster center follows a Brownian motion in a tilted cosine potential with the same wavelength and drag force, but with a reduced amplitude $(U_0/n)\sin(n\pi\sigma)/\sin(\pi\sigma)$ of the external potential \cite{Lips/etal:2021}. Due to the factor $1/n$, this amplitude is of the order of $k_{\rm B}T=1$ for cluster sizes $n\ge 4$ and $U_0=6$. As the forces 
given by the external potential alternate in sign, we may neglect them compared to the constant drag force, i.e., we roughly estimate the cluster velocity to equal $f$. This gives $t_\sigma\simeq(1-\sigma)/f$ and we hence obtain
\begin{equation}
v_{\rm d}(\sigma)\simeq \alpha\,\frac{f}{1-\sigma}\,.
\label{eq:vd}
\end{equation}
The proportionality factor $\alpha$ should be of the order of one and only weakly dependent on $\sigma$.
In fact, when choosing $\alpha=1.35$, we obtain a quite good agreement of the estimate in Eq.~(\ref{eq:vd}) 
with the simulated data, see Fig.~\ref{fig:defect-velocity}.

Equation~(\ref{eq:vd}) would predict that the defect velocity becomes arbitrary large in the limit $\sigma\to1$, i.e.\ when the particle diameter approaches the wavelength of the cosine potential.  However, as explained above, a defect formation requires the formation of a hole as an initial step. A hole can only form for $\sigma<(L-1)/L$, i.e.\ the limit $\sigma\to1$ cannot be taken in a finite system. The thermodynamic limit $L\to\infty$ is discussed in the next section.

\begin{figure}[t!]
\center{\includegraphics[width=0.6\textwidth]{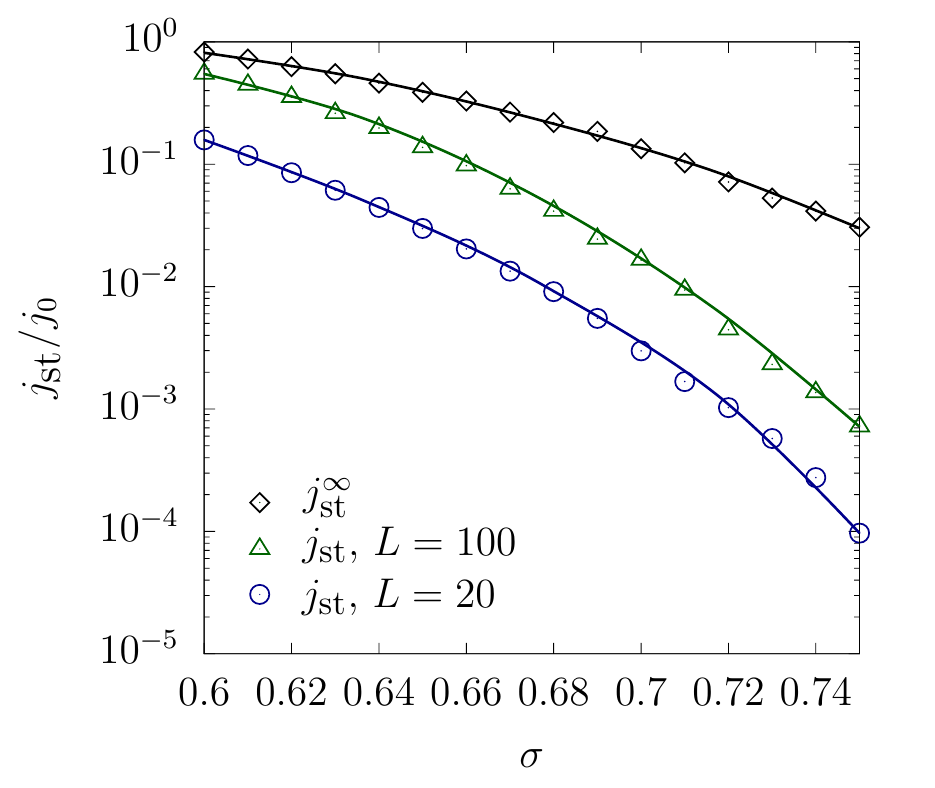}}
\caption{Steady-state current $j_{\rm st}$ as a function of $\sigma$ for different system sizes $L$ compared to its value $j_{\rm st}^\infty$ in the limit $L\to\infty$. As in Fig.~\ref{fig:j-sigma}, the currents are normalized with respect to the current $j_0$ of noninteracting particles. The lines are drawn as guide for the eyes.}
\label{fig:j-sigma_different_L}
\end{figure}

\section{Defect-mediated current}
\label{sec:defect-mediated-current}
Knowing the defect velocitiy $v_{\rm d}$ and generation rate $\lambda_{\rm{gen}}$, we can calculate the defect-mediated current for large systems (thermodynamic limit of infinite system size), where defects are constantly generated and annihilated. Annihilation occurs after some lifetime if a defect encounters a hole. Accordingly, the mean lifetime $\tau$ is equal to $\tau=d_h/v_{\rm d}$, where $d_h$ is the mean distance between holes. This is given by $d_h=1/\varrho_h$, where $\varrho_h$ is the number density of holes, which is also the number density of defects, $\varrho=\varrho_h$. This varies in time as
\begin{eqnarray}\label{eq:pdr_lattice}
\frac{{\rm d} \varrho}{{\rm d}t}=\lambda_{\rm{gen}}-\frac{\varrho}{\tau}\,.
\end{eqnarray}
In the steady state, we obtain $\varrho_{\rm st}=\lambda_{\rm{gen}}\tau=\lambda_{\rm gen}/\varrho_{\rm st} v_{\rm d}$, i.e.,
$\varrho_{\rm st}=(\lambda_{\rm{gen}}/v_{\rm d})^{1/2}$. The defect-mediated current in the steady state then is
\begin{equation}
j_{\rm st}^\infty=\varrho_{\rm st}\, v_{\rm d}=\sqrt{\lambda_{\rm gen}v_{\rm d}}\,.
\label{eq:jst-theory}
\end{equation}

As shown in Fig.~\ref{fig:j-sigma_different_L}, the results predicted by 
Eq.~(\ref{eq:jst-theory}) in the limit of infinite system size do not agree with the simulated data for $L=20$ and $L=100$.
In particular, the decrease of $j_{\rm st}^\infty$ with $\sigma$ is much weaker and it does
not resemble an apparent jamming transition.

The reason for this mismatch is that the current becomes strongly affected by the finite size of system, when $L$ becomes smaller than a distance $L_\sigma$ proportional to the mean distance $1/\varrho_{\rm st}$ between the 
defects in the infinite system. In the following, we set $L_\sigma=(v_{\rm d}/\lambda_{\rm gen})^{1/2}$. 
Accordingly, the length $L_\sigma$ is expected to be very large
due to the small rate for the rare events of defect generation. Unfortunately, in the Brownian dynamics simulations of hard spheres
at filling factor one and large $\sigma\gtrsim0.5$, one cannot explore the dynamics at large lengths $L$ in a reasonable computing 
time due to the large number of collisions between the spheres. These collisions must be treated carefully in the algorithm 
\cite{Scala/etal:2007}, in particular regarding the choice of a proper integration time step and the treatment of multiple collisions 
during one time step \footnote{For example, for the small system size $L=20$,  the 
CPU time to obtain a single value $j_{\rm st}$ for $\sigma=0.75$
is about 230~hours on a 16-core Intel Xeon Nehalem 2.66~GHz processor.}.

\begin{figure}[t!]
\center{\includegraphics[width=0.4\textwidth]{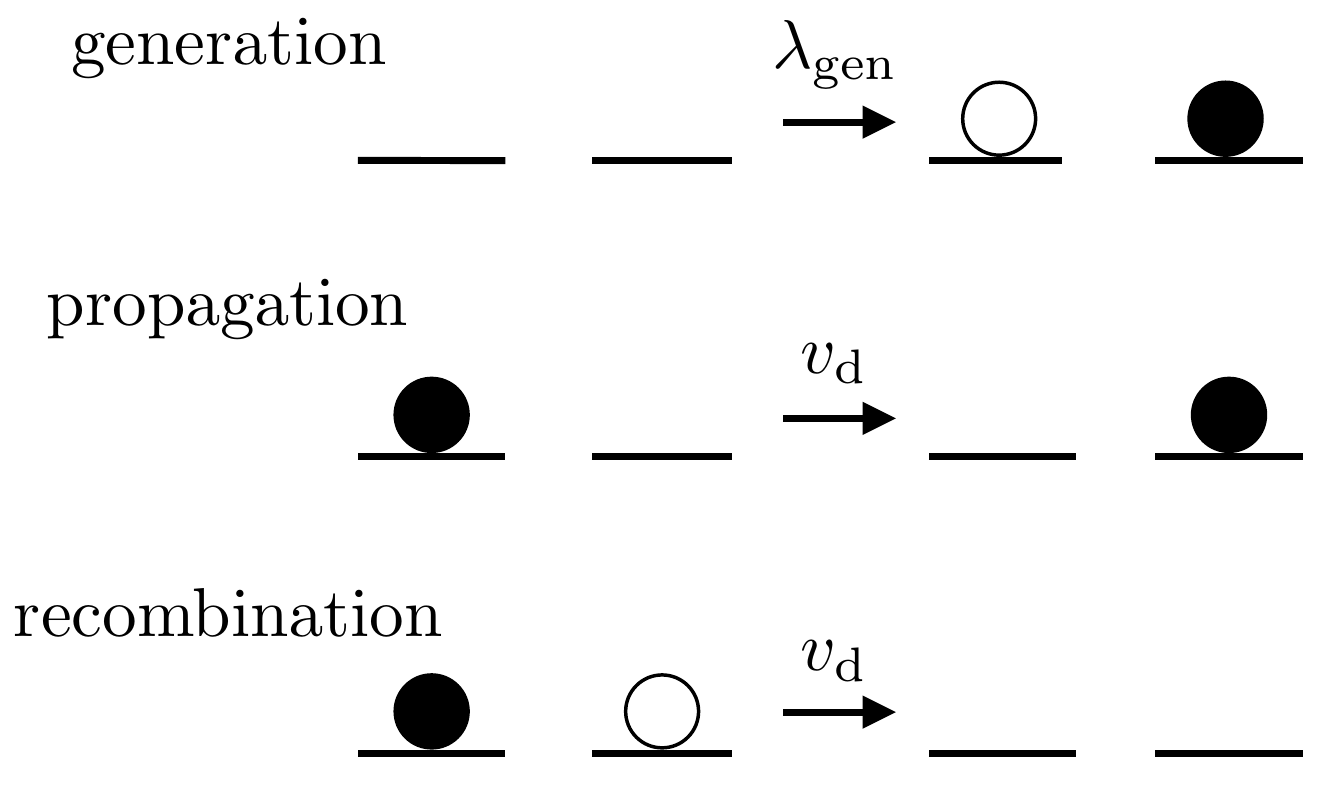}}
\caption{Possible elementary processes in the coarse-grained lattice model.}
\label{fig:illustration-tasep}
\end{figure}

Hence, we thought of overcoming this problem by developing a coarse-grained lattice model of the defect dynamics. 
In this model, illustrated in Fig.~\ref{fig:illustration-tasep}, particles represent defects and holes represent unoccupied potential wells. A lattice site can be either occupied by a hole or a particle, or it can be empty. Double occupancies of lattice sites by holes and particles are forbidden. The holes cannot move, which reflects that their motion is much slower than that of the defects. The particles can jump to a right nearest-neighbor site with a rate $v_{\rm d}$ if this site is empty or occupied by a hole. In the latter case, the particle and hole recombine and the site becomes empty. Holes and particles are generated in pairs at nearest-neighbor empty sites with the rate $\lambda_{\rm gen}$. This model
can be considered as a variant of the two-species totally asymmetric simple exclusion process \cite{Evans/etal:1995, Bonnin/etal:2022}, 
where the particle-hole generation and recombination are specific features here.

\begin{figure}[t!]
\center{
\includegraphics[width=\textwidth]{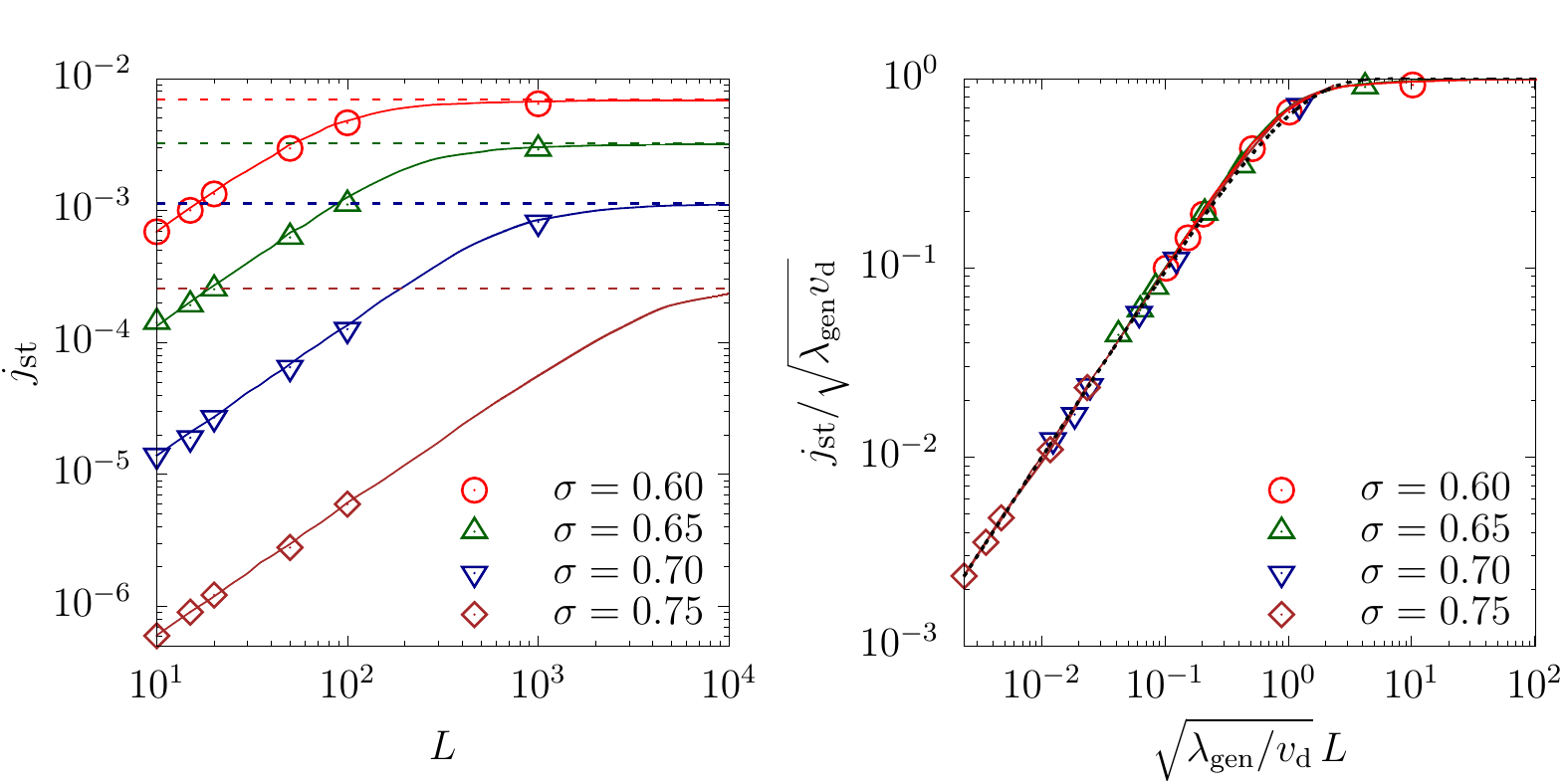}}
\caption{(a) Dependence of the steady-state current $j_{\rm st}$ on the system size $L$ for various particle diameters $\sigma$. 
The solid lines are from kinetic Monte 
Carlo simulations of the coarse-grained lattice model, cf.\ Fig.~\ref{fig:illustration-tasep}. They agree well 
with the results from Brownian dynamics simulations (symbols) and
for large $L$ approach the theoretically predicted value $j_{\rm st}^\infty$ given in
Eq.~(\ref{eq:jst-theory}) [indicated by the dashed horizontal lines]. (b)~Scaled current vs.\ scaled system size demonstrating the data collapse onto a common master curve 
(solid line).
The master curve corresponds to the scaling function $G(.)$ in Eq.~(\ref{eq:j-scaling}) and the dashed line represents the 
approximation~(\ref{eq:G-approximation}).}
\label{fig:j_L}
\end{figure}

The current in the lattice model is obtained by counting the number of particle jumps per time. 
In Fig.~\ref{fig:j_L} it is shown how the currents vary with $L$ for different $\sigma$. 
For $L\le 10^3$ they agree well with the currents of the BASEP, which confirms the proposed mechanism of defect-mediated currents. The lattice model thus allows to uncover the full dependence on $L$. 
For $L\lesssim L_\sigma=1/\varrho_{\rm st}$, the current is limited by the low total rate $\lambda_{\rm gen}L$ of generating defects,
hence $j_{\rm st}\simeq \lambda_{\rm gen}L$. For $L\gg1/\varrho_{\rm st}$, the current  approaches $j_{\rm st}^\infty$
given in Eq.~(\ref{eq:jst-theory}).

The dependence of $j_{\rm st}=j_{\rm st}(\sigma,L)$ on both $\sigma$ and $L$ can be described by the scaling form
$j_{\rm st}(\sigma,L)=j_{\rm st}^\infty\, G(L/L_\sigma)$, or 
\begin{equation}
j_{\rm st}(\sigma,L)=\sqrt{\lambda_{\rm gen}v_{\rm d}}\; G\Biggl(\sqrt{\frac{\lambda_{\rm gen}}{v_{\rm d}}} L\Biggr)\,.
\label{eq:j-scaling}
\end{equation}
This scaling is demonstrated in Fig.~\ref{fig:j_L}(b), where the simulated data collapse onto the master curve $G(x)$. 
Because $j_{\rm st}\simeq \lambda_{\rm gen}L$ for small $L$, we must have $G(x)/x\to1$ for $x\to0$. For $x\to\infty$,
$G(x)\to1$. A simple approximate form of the scaling function is
\begin{equation}
G(x)\simeq 1-e^{-x}\,.
\label{eq:G-approximation}
\end{equation}
This is shown as the dashed line in Fig.~\ref{fig:j_L}(b).

\section{Summary and Conclusions}
We have shown that an apparent jamming transition occurs in the driven Brownian motion of hard spheres through a periodic 
potential under a constant drag force $f$, where the particle number equals the number of potential wells (filling factor one).
This jamming transition manifests itself in steady-state currents $j_{\rm st}$
decreasing by several orders of magnitude if the particle diameter $\sigma$ is 
increased beyond 0.6 wavelengths of the periodic potential. The current in the apparently jammed states is mediated by defects 
that are collective thermal excitations. 

These defects are appearing as rare events. Determining their properties in many-particle Brownian dynamics simulations is therefore a challenging task. We have obtained reliable results for currents and defect generation rates
by direct simulation for system sizes $L$ up to $10^3$~wavelengths. 
For evaluating defect velocities $v_{\rm d}$, the largest possible $L$ were even much smaller ($L\lesssim 50$).
To overcome difficulties with the very rare defect generation, we have applied a particle insertion method 
for obtaining $v_{\rm d}$ much more efficiently. 
We have designed a coarse-grained lattice 
model to fully uncover the current behavior as a function of system size.

To describe the defect generation, we have developed a transition state theory. It 
predicts defect generation rates $\lambda_{\rm gen}$ which decay faster than exponentially with $\sigma$ 
and are in excellent quantitative 
agreement with simulations. The number of particles forming the transition state follows from
a maximum free energy barrier principle. 

The velocity $v_{\rm d}$ of the defects is hundred times larger than that of a single (noninteracting) 
particle dragged through the periodic potential, and even several times larger than that of a single dragged particle in the absence 
of the periodic potential. We have explained this ultrafast defect propagation by considering a coherent-like motion of 
particles within a defect cluster. The potential amplitude for the cluster dynamics is effectively reduced. By further taking into 
account geometric relations,
we have obtained the approximate expression $v_{\rm d}\propto f/(1-\sigma)$.

By combining the results for the defect velocity and generation rate, we have derived a scaling law for describing the 
dependence of the currents on $\sigma$ and the system size $L$. There are two scaling regimes of small and large $L$, which 
are separated by the mean distance $L_\sigma$ between the defects in the thermodynamic limit, i.e.,
$L_\sigma=1/\varrho_{\rm st}$, where $\varrho_{\rm st}$ is the number density of defects.
For small system size $L\lesssim L_\sigma$, the current is governed by the defect generation rate solely, 
$j_{\rm st}\simeq\lambda_{\rm gen}L$. Thus the apparent jamming transition is caused by the strong 
decrease of $\lambda_{\rm gen}$ with $\sigma$. For large $L$, 
$j_{\rm st}\simeq \sqrt{\lambda_{\rm gen}v_{\rm d}}$, meaning that the decrease of $j_{\rm st}$ 
with $\sigma$ is much weaker than in the scaling regime of small $L$. 

It is remarkable that the complex rare event dynamics of the defects and their connection to currents could be accounted for
by the theoretical concepts introduced above. We believe that the method of identifying many-particle transition states and of 
calculating their formation rates can be applied to describe the collective motion of current carriers in various crowded systems.
The dependence of the defect velocity on the particle size appears to be a typical feature of cluster dynamics.
Similarly, the scaling behavior of currents with respect to system size seems to be a generic feature for charge transport in 
one-dimensional conductors. 

Nowadays, many experimental techniques are capable to control and track the motion of microparticles in crowded  environments \cite{Bohlein/etal:2012, Vanossi/etal:2012, Tierno:2014, Arzola/etal:2017, Pagliara/etal:2013, Pagliara/etal:2014, Juniper/etal:2015, Locatelli/etal:2016, Skaug/etal:2018, Schwemmer/etal:2018, Zemanek/etal:2019, Battat/etal:2022}.
The apparent jamming transition should be well accessible by measurements because it occurs already for particle diameters 
significantly smaller than the wavelength of the periodic potential. For experiments with colloidal particles confined to
channels of optical or magnetic traps, the number of particles is about 10-100. In that range, the apparent jamming transition
is particularly pronounced. Moreover, the particle insertion method can be used to study defect 
propagation in a controlled way and to overcome possible problems with the observation of rare events.
Such experiments can critically test the theoretical predictions and possibly uncover new unexplored features of 
collective excitations in dense many-particle systems.

\noindent\textbf{Acknowledgements}\\
Financial support by the Czech Science Foundation (Project No.\ 20-24748J) and the Deutsche Forschungsgemeinschaft (Project No.\ 432123484) is gratefully acknowledged. A part of computational resources was supplied by the project ``e-Infrastruktura CZ'' (e-INFRA CZ LM2018140) supported by the Ministry of Education, Youth and Sports of the Czech~Republic.

%

\end{document}